\documentclass[prd,twocolumn]{revtex4}
\usepackage{bm}
\usepackage{amssymb}
\usepackage{graphicx}
\begin{document}
\title{The gravitational self-force} 
\author{Eric Poisson}
\affiliation{Department of Physics, University of Guelph, Guelph,
Ontario, Canada N1G 2W1}
\date{October 24, 2004} 
\begin{abstract}
The self-force describes the effect of a particle's own gravitational
field on its motion. While the motion is geodesic in the test-mass
limit, it is accelerated to first order in the particle's mass. In
this contribution I review the foundations of the self-force, and show
how the motion of a small black hole can be determined by matched
asymptotic expansions of a perturbed metric. I next consider the case
of a point mass, and show that while the retarded field is singular on
the world line, it can be unambiguously decomposed into a singular
piece that exerts no force, and a smooth remainder that is responsible
for the acceleration. I also describe the recent efforts, by a number
of workers, to compute the self-force in the case of a small body
moving in the field of a much more massive black hole.    

The motivation for this work is provided in part by the Laser
Interferometer Space Antenna, which will be sensitive to low-frequency 
gravitational waves. Among the sources for this detector is the motion 
of small compact objects around massive (galactic) black holes. To
calculate the waves emitted by such systems requires a detailed 
understanding of the motion, beyond the test-mass approximation. 

This article is based on a plenary lecture presented at the 17th
International Conference on General Relativity and Gravitation, which
took place in July, 2004 in Dublin, Ireland.   
\end{abstract}
\maketitle

\section{Introduction: The Capra scientific mandate} 
\label{i}

This contribution describes how a body of mass $m$, supposed to be 
``small'', moves in spacetime. In the test-mass approximation it is
known that the body moves on a geodesic of a background spacetime
whose metric $g_{\alpha\beta}$ does not depend on $m$. As $m$
increases (but kept small) the test-mass description is no longer
adequate. One might then say that the motion is still geodesic, but in
a spacetime whose metric is perturbed with respect to the background
metric. Alternatively, one might say that the motion is accelerated in
the background spacetime. This is the point of view I shall take in
this contribution: The body's motion will be described in the
original (unperturbed) spacetime, and the gravitational influence of
the body will be incorporated in its acceleration. The body will be
said to move under the influence of its gravitational self-force. 

The work described here was previously reviewed in a very long article
published in {\it Living Reviews in Relativity} \cite{poisson:04b}. An
abridged version of this article appeared in {\it Classical and
Quantum Gravity} \cite{poisson:04c}. This contribution focuses on the
highlights and presents the ``big picture''. The reader is warned that
my presentation will be sketchy and incomplete, and is referred to the
review articles for additional details. The scientific objectives
behind the work described here have been pursued by a number of
people; I like to refer to these objectives as the ``Capra scientific
mandate'' and to these people as the ``Capra posse''. 

The Capra scientific mandate includes three main objectives. The first
is {\it to formulate the equations of motion of a small body in a
specified background spacetime, beyond the test-mass approximation.}
This first step was solved in 1997, first by Mino, Sasaki, and Tanaka
\cite{mino-etal:97}, and then by Quinn and Wald
\cite{quinn-wald:97}. The equations of motion are now known as the
MiSaTaQuWa equations; I will sketch their derivation in this
contribution. The second objective is {\it to concretely describe the
motion of the small body in situations of astrophysical interest,
including generic orbits of a Kerr black hole}. There has been much
recent progress on this front, and I will describe some of the issues
involved in this contribution. The third objective is {\it to properly
incorporate the equations of motion into a wave-generation formalism.}
This final objective is the most important, as the ultimate goal of
this enterprise is to make detailed predictions toward eventual
gravitational-wave measurements. This is the holy grail of the Capra
program, and it has so far proved elusive. I will present a tentative
outline of future work in the last section of this contribution. 

The work reviewed in this contribution was shaped by a series of
annual meetings named after the late movie director Frank Capra. The
first of these meetings took place in 1998 and was held at Capra's
ranch in Southern California; the ranch now belongs to Caltech,
Capra's alma mater. Subsequent meetings were held in Dublin, Pasadena,
Potsdam, State College PA, Kyoto, and Brownsville. 

Members of the Capra posse include Paul Anderson, Warren Anderson,
Leor Barack, Patrick Brady, Lior Burko, Manuella Campanelli, Steve
Detweiler, Eanna Flanagan, Costas Glempedakis, Abraham Harte, Wataru
Hikida, Bei Lok Hu, Scott Hughes, Sanjay Jhingan, Dong-Hoon Kim,
Carlos Lousto, Eirini Messaritaki, Yasushi Mino, Hiroyuki Nakano, Amos
Ori, Ted Quinn, Eran Rosenthal, Norichika Sago, Misao Sasaki, Takahiro
Tanaka, Bob Wald, Bernard Whiting, and Alan Wiseman.  

\section{Astrophysical context} 
\label{ii}

The motivation for the Capra program comes largely from the fact that 
solar-mass compact bodies moving around massive black holes have been
identified as one of the promising sources of gravitational waves for
the space-based interferometric detector LISA (Laser Interferometer
Space Antenna). (The case for this identification is made in the
contributions by Sterl Phinney and Sir Martin Rees.) These systems
involve highly eccentric, nonequatorial, and relativistic orbits
around rapidly rotating black holes. The waves produced by these
orbits will be rich in information concerning the strongest
gravitational fields in the Universe, and this information will be
extractable from the LISA data stream. The extraction, however, will
depend on sophisticated data-analysis strategies that will rely on a
detailed and accurate modeling of the source. This modeling involves 
formulating the equations of motion for the small body in the field of
the rotating black hole, {\it in a small-mass-ratio approximation that
goes beyond the test-mass description}. And it involves a consistent 
incorporation of these equations of motion into a wave-generation
formalism. In short, the extraction of this wealth of information
relies on the successful completion of the Capra program.       

The finite-mass corrections to the orbiting body's motion are
important. Let $m$ be the mass of the orbiting body, $M$ the mass of
the central black hole, and suppose that $m/M \ll 1$. For
concreteness, assume that the orbiting body is a $10\ M_\odot$ black
hole and that the central black hole has a mass of $10^6\
M_\odot$. Then $m/M = 10^{-5}$ is the order of magnitude of the
correction to the equations of motion relative to the test-mass
description. Simultaneously, $M/m = 10^5$ is the order of magnitude of
the total number of wave cycles that will be received during a year's
worth of LISA observation. This simplistic estimate illustrates that
while the corrections to the equations of motion are small, in the
course of a year they can accumulate and contribute a significant
number of wave cycles.    

Corrections to the equations of motion must incorporate both
conservative and dissipative effects. Finite-mass corrections that are
conservative in nature are familiar from Newtonian and post-Newtonian
theory, and they occur also in a strong-gravity situation. These can
accumulate over time. Imagine, for example, an eccentric orbit that
undergoes periastron advance. A finite-mass correction to this effect
will cause a steady drift in the phasing of the orbit, and this will 
directly be reflected in the phasing of the gravitational waves; after 
$10^5$ orbital cycles the $10^{-5}$ correction will have grown into a
sizable effect. 

Dissipative effects, on the other hand, do not occur in Newtonian
gravity, but they are familiar in post-Newtonian theory; they are also
present in a strong-gravity situation. Dissipation is associated with
the radiative loss of energy and angular momentum by the orbital
system, and the resulting corrections to the motion also accumulate
over time; this translates into a steady drift of the
gravitational-wave signal in the LISA frequency band. These are
finite-mass corrections, because the test-mass description makes no
room for gravitational radiation and radiation reaction.  

From a more theoretical point of view, the appeal of this work comes
largely from the fact that while the motion of self-gravitating bodies
has been studied extensively in the context of post-Newtonian theory
(see, for example, Refs.~\cite{damour:87, blanchet:02}), very little
is known in the case of strong fields and fast motions. To the
relativist working in this area, this problem is irresistible: We 
have strong gravity, fast motion, and a smallness parameter in the
form of $m/M$. We have a cool problem that can be solved by standard
perturbative techniques.   

\section{Motion of a black hole} 
\label{iii}

Let me restate the problem in its most general form: A body of mass
$m$ moves in an arbitrary (but empty of matter) spacetime whose radius
of curvature (in the body's neighbourhood) is ${\cal R}$; what is the
description of its motion when $m \ll {\cal R}$? This formulation of
the problem is more general than the two-body version stated
previously. When the small body is a member of a binary system, and is
at a distance $b$ from another body of mass $M$, then ${\cal R} \sim 
\sqrt{b^3/(M+m)}$ and 
\[
\frac{m}{{\cal R}} \sim \frac{m}{M+m} v^3,
\]
where $v = \sqrt{(M+m)/b}$. For relativistic motion ($v \sim 1$) this
is small whenever $m/M \ll 1$.      

The clean separation of scales allows us to idealize the motion as
following a world line in a spacetime whose metric $g_{\alpha\beta}$
will be specified below. While the region of spacetime occupied by the
body is truly a world tube of finite extension, on a scale ${\cal R}$
--- the only scale of relevance in the background spacetime --- this
extension is so small that little is lost by making this
idealization. The body shall then follow a world line $\gamma$ that
will be described by parametric relations $z^\mu(\tau)$, in which
$\tau$ is proper time measured in the metric $g_{\alpha\beta}$. We
wish to determine this world line. It is understood that the world
line loses all significance when the neighbourhood of the body is
examined on the fine scale $m$; on this scale the finite extension of
the world tube is fully revealed. 

It is desirable to choose the body to have an internal structure that
is as simple as possible. One thus eliminates an important source of
technical complications, but one still feels confident that the
resulting equations of motion will apply, to a good degree of
accuracy, to a body of arbitrary structure. (This ``effacement
property'' of the internal structure is well established in
post-Newtonian theory \cite{damour:87}.) This attitude often leads the 
researcher to assume that the body is a point particle. We shall
refrain from doing so at this stage, but we will come back to this
description in Sec.~\ref{iv}. We shall instead choose the body to have
the simplest structure compatible with the laws of general relativity:
it shall be a nonrotating black hole. This was one of the starting
assumptions of Mino, Sasaki, and Tanaka \cite{mino-etal:97} (see also 
Ref.~\cite{poisson:04a}).  

The motion of the black hole is determined by {\it matched asymptotic
expansions}, a powerful technique that has known many useful
applications in general relativity (see, for example,
Refs.~\cite{thorne-hartle:85, death:96}). In this approach the
metric of the black hole perturbed by the tidal gravitational field
supplied by the external universe is matched to the metric of the
external universe perturbed by the moving black hole. Demanding
that this metric be a valid solution to the Einstein field equations
determines the motion without additional input.  
 
The method of matched asymptotic expansions relies on the existence of
(i) an {\it internal zone} in which gravity is dominated by the black
hole, (ii) an {\it external zone} in which gravity is dominated by the
conditions in the external universe, and (iii) an overlapping region
known as the {\it buffer zone}, in which the black hole and the
external universe have comparable influence. Let $r$ be a meaningful
measure of spatial distance from the black hole. Then the internal
zone is the region of spacetime in which $r < r_i = \mbox{constant}$,
where $r_i \ll {\cal R}$; thus $r/{\cal R}$ is small throughout the
internal zone and the tidal influence of the external universe is
small. The external zone, on the other hand, is the region of
spacetime in which $r > r_e = \mbox{constant}$, where $r_e \gg m$;
thus $m/r$ is small throughout the external zone and the gravitational
effects of the black hole are small. Finally, the buffer zone is the
region of spacetime in which $r_e < r < r_i$. This region exists
because $m \ll {\cal R}$, and $r/{\cal R}$ and $m/r$ are
simultaneously small in the buffer zone. 

An expansion of the metric in the internal zone has the schematic form 
\begin{equation}
{\sf g} = g(\mbox{Schwarzschild}) + O(r^2/{\cal R}^2) + \cdots,        
\label{1}
\end{equation}
where $g(\mbox{Schwarzschild})$ is the metric of a nonrotating black
hole in isolation, and $O(r^2/{\cal R}^2)$ is the tidal field 
supplied by the external universe. (Terms that scale as $r/{\cal R}$
have been removed by a coordinate transformation.) On the
other hand, an expansion of the metric in the external zone has the
schematic form 
\begin{equation}
{\sf g} = g(\mbox{background}) + O(m/r) + \cdots, 
\label{2}
\end{equation} 
where $g(\mbox{background})$ is the metric of the external universe in
the absence of the black hole, and $O(m/r)$ is the perturbation
produced by the moving black hole. The key idea of matched asymptotic
expansions is that the expansions of Eqs.~(\ref{1}) and (\ref{2}) are
both valid in the buffer zone, and that both forms of ${\sf g}$
represent the same metric (up to a coordinate
transformation). Performing the matching returns the black hole's
equations of motion.   

\subsection*{External zone} 

To flesh out these ideas let us first examine the situation in the
external zone. Let $g_{\alpha\beta}$ stand for $g(\mbox{background})$,
the metric of the external universe in the absence of the small black
hole. (In the astrophysical context of Sec.~\ref{ii} this would be the
metric of the massive black hole, in isolation.) Let $\gamma$ be a
fiducial world line in this background spacetime (later to be
identified with the small hole's world line), and express the metric
in normal coordinates centered on $\gamma$. It will have the form
of $g = \eta + O(r/{\cal R}) + O(r^2/{\cal R}^2) + \cdots$, where
$\eta$ is the metric of flat spacetime. The normal coordinates do not
extend beyond $r = {\cal R}$ and are therefore restricted to the
buffer zone.  

More specifically, let the normal coordinates be the retarded
coordinates $(u,x^a = r\Omega^a)$, such that $u$ is constant on each
future light cone centered on $\gamma$ (and $u$ is equal to proper
time on the world line), $r$ is an affine parameter on the cone's null
generators, and $\Omega^a \equiv x^a/r$ is constant on each
generator. Then the time-time component of the background metric takes
the form  
\begin{equation}
g_{uu} = -\bigl(1 + 2 r a_a \Omega^a + r^2 {\cal E}_{ab} \Omega^a 
\Omega^b \bigr) + O(r^3/{\cal R}^3), 
\label{3}
\end{equation} 
where $a_a(u)$ is the acceleration vector of the world line, and
${\cal E}_{ab}(u) = C_{uaub} = O(1/{\cal R}^2)$ are the electric
components of the Weyl tensor evaluated on $\gamma$. (This
three-tensor is symmetric and tracefree.) Our main goal is
to eventually determine $a_a$; we can anticipate that $a_a = 
O(m/{\cal R}^2)$ since the motion must be geodesic in the test-mass
limit. 

Let $h_{\alpha\beta}$ be the perturbation produced by the moving black
hole, so that the full metric is ${\sf g}_{\alpha\beta} =
g_{\alpha\beta} + h_{\alpha\beta}$. A standard technique in
perturbation theory is to introduce the trace-reversed potentials 
\begin{equation}
\psi_{\alpha\beta} = h_{\alpha\beta} - \frac{1}{2} g_{\alpha\beta}
\bigl( g^{\gamma\delta} h_{\gamma\delta} \bigr) 
\label{4}
\end{equation}
and to impose the Lorenz gauge condition 
\begin{equation} 
\psi^{\alpha\beta}_{\ \ \ ;\beta} = 0. 
\label{5}
\end{equation} 
Here and below, all indices are manipulated with the background
metric, and covariant differentiation is taken to be compatible with 
this metric. 

In the external zone the perturbation produced by the black hole
cannot be distinguished from one produced by a point particle of mass
$m$ moving on $\gamma$, and $\psi_{\alpha\beta}$ must be a solution to
the wave equation  
\begin{equation} 
\Box \psi^{\alpha\beta} + 2 R^{\alpha\ \beta}_{\ \gamma\ \delta} 
\psi^{\gamma\delta} = -16\pi T^{\alpha\beta},        
\label{6}
\end{equation}
where $T^{\alpha\beta}$ is the particle's stress-energy tensor. This
equation can be solved by means of a retarded Green's function, 
\begin{equation} 
G^{\alpha\beta}_{\ \ \gamma'\delta'}(x,x') = 
U^{\alpha\beta}_{\ \ \gamma'\delta'}(x,x') \delta(\sigma) 
+ V^{\alpha\beta}_{\ \ \gamma'\delta'}(x,x') \theta(-\sigma), 
\label{7}
\end{equation}
where $\sigma(x,x')$ is half the squared geodesic distance between the
points $x$ and $x'$, and $U^{\alpha\beta}_{\ \ \gamma'\delta'}(x,x')$, 
$V^{\alpha\beta}_{\ \ \gamma'\delta'}(x,x')$ are smooth
bitensors. The Green's function is decomposed into a
singular ``light-cone part'' that has support on $\sigma = 0$ only,
and a smooth ``tail part'' that has support on $\sigma < 0$ (so that
$x$ is in the chronological future of $x'$). (It should be noted that
this representation of the Green's function is valid only when $x$ is
in the normal convex neighbourhood of $x'$. In the sequel I will
simplify expressions by pretending that the decomposition holds
globally. The reader is referred to LRR \cite{poisson:04b} for all the
fine print.)   

The solution to Eq.~(\ref{6}) is 
\begin{equation} 
\psi^{\alpha\beta}(x) = \frac{4m}{r} U^{\alpha\beta}_{\ \
\gamma'\delta'}(x,x') u^{\gamma'} u^{\delta'} 
+ \psi^{\alpha\beta}_{\rm tail}(x) 
\label{8}
\end{equation} 
where $x' \equiv z^\mu(u)$ is the point on the world line that is
linked to $x$ by a null geodesic, $u$ the value of the proper-time
parameter at this retarded point, $u^{\gamma'}$ the four-velocity at 
the retarded point, and   
\begin{equation} 
\psi^{\alpha\beta}_{\rm tail}(x) = 4m \int_{-\infty}^u
V^{\alpha\beta}_{\ \ \mu'\nu'}(x,z') u^{\mu'} u^{\nu'}\, d\tau' 
\label{9}
\end{equation} 
is the ``tail'' term. Here $z' \equiv z(\tau')$ stands for an
arbitrary position on the world line, and the potentials of
Eq.~(\ref{8}) depend on the particle's entire history prior to the
retarded point $\tau = u(x)$. 

Inverting Eq.~(\ref{4}) and expressing the results in the retarded
coordinates $(u,r\Omega^a)$ returns 
\begin{eqnarray} 
{\sf g}_{uu} &=& -1 - r^2 {\cal E}_{ab} \Omega^a \Omega^b 
+ O(r^3/{\cal R}^3)  
\nonumber \\ & & \mbox{} 
+ \frac{2m}{r} + h^{\rm tail}_{uu} + r \bigl( 2m {\cal E}_{ab}
\Omega^a \Omega^b - 2 a_a \Omega^a 
\nonumber \\ & & \mbox{} 
+ h^{\rm tail}_{uuu} + h^{\rm tail}_{uua} \Omega^a \bigr) 
+ O(mr^2/{\cal R}^3)
\label{10} 
\end{eqnarray}  
for the time-time component of the perturbed metric 
${\sf g}_{\alpha\beta} = g_{\alpha\beta} + h_{\alpha\beta}$. The
fields $h^{\rm tail}_{\alpha\beta}$ are obtained from 
$\psi^{\rm tail}_{\alpha\beta}$ by trace reversal, and 
$h^{\rm tail}_{\alpha\beta\gamma} = \nabla_\gamma 
h^{\rm tail}_{\alpha\beta}$. 

Equation (\ref{10}) gives the metric of the external universe
perturbed by a black hole moving on a world line $\gamma$ with an 
(as yet undetermined) acceleration $a_a(u)$. It is noteworthy that 
this metric appears to be singular when $r \to 0$. The metric,
however, is valid only in the external zone $r > r_e \gg m$, and the 
limit $r \to 0$ is unattainable. The world line $\gamma$ is outside
the metric's domain of validity.  

\subsection*{Internal zone} 

To obtain the metric of a nonrotating black hole that is slightly
distorted by a tidal gravitational field is a standard application of
black-hole perturbation theory. To leading order in the perturbation,
which scales as $1/{\cal R}^2$, it is sufficient to integrate a set of 
time-independent perturbation equations, because each time derivative
comes with an extra factor of $1/{\cal R}$, the inverse time
scale associated with changes in the external universe. The
perturbation must be well behaved on the hole's event horizon, and the 
asymptotic conditions for $r \gg 2m$ are such that the perturbation
must behave as $-r^2 {\cal E}_{ab} \Omega^a \Omega^b$, which has a
quadrupolar form. These observations make solving the perturbation
equations a very straightforward task.   

In a set of internal light-cone coordinates $(\bar{u},\bar{x}^a =
\bar{r} \bar{\Omega}^a)$ the time-time component of the perturbed
metric reads  
\begin{equation} 
{\sf g}_{\bar{u}\bar{u}} = -f - \bar{r}^2 f^2 {\cal E}_{ab} 
\bar{\Omega}^a \bar{\Omega}^b + O(\bar{r}^3/{\cal R}^3), 
\label{11}
\end{equation} 
where $f = 1 - 2m/\bar{r}$ and ${\cal E}_{ab}(\bar{u})$ is the 
tidal gravitational field supplied by the external universe. In the
limit $m/\bar{r} \to 0$ (keeping $\bar{r}/{\cal R}$ fixed)
Eq.~(\ref{11}) becomes the metric of the external universe, expressed 
in coordinates for which $a_a = 0$. In the limit 
$\bar{r}/{\cal R} \to 0$ (keeping $m/\bar{r}$ fixed) Eq.~(\ref{11})
becomes the metric of a Schwarzschild black hole expressed in
Eddington-Finkelstein coordinates. For small values of 
$\bar{r}/{\cal R}$ and arbitrary values of $m/\bar{r}$ Eq.~(\ref{11})
describes a tidally distorted black hole. The metric of Eq.~(\ref{11})
is valid everywhere in the internal zone, where 
$r < r_i \ll {\cal R}$.   

\subsection*{Matching} 

Equation (\ref{10}) gives the spacetime metric in the external zone,
and the requirement that $r/{\cal R} \ll 1$ implies that the metric is 
in fact restricted to the buffer zone. Equation (\ref{11}), on the
other hand, gives the spacetime metric in the internal zone, and its
specialization to $m/\bar{r} \ll 1$ implies also a restriction to the
buffer zone. Since both metrics describe the geometry of the same
region of the same spacetime, they must be related by a coordinate
transformation. 

The transformation from the external coordinates $(u,x^a)$ to the
internal coordinates $(\bar{u},\bar{x}^a)$ is given in part by 
\begin{eqnarray} 
\bar{u} &=& u - 2m \ln r 
- \frac{1}{2} \int^{u} h^{\rm tail}_{uu}\, du  
- \frac{1}{2} r \Bigl[ h^{\rm tail}_{uu} 
+ 2 h^{\rm tail}_{ua} \Omega^{a} 
\nonumber \\ & & \mbox{} 
+ h^{\rm tail}_{ab} \Omega^{a} \Omega^{b} \Bigr] 
- \frac{1}{4} r^{2} \Bigl[ 
h^{\rm tail}_{uuu} + \bigl(h^{\rm tail}_{uua} 
+ 2 h^{\rm tail}_{uau} \bigr) \Omega^{a} 
\nonumber \\ & & \mbox{} 
+ \bigl( h^{\rm tail}_{abu} + 2 h^{\rm tail}_{uab} \bigr)    
\Omega^{a} \Omega^{b} 
+ h^{\rm tail}_{abc} \Omega^{a} \Omega^{b} \Omega^{c} \Bigr] 
\nonumber \\ & & \mbox{} 
+ O(mr^3/{\cal R}^3). 
\label{12}
\end{eqnarray} 
Applying this transformation to Eq.~(\ref{10}) gives  
\begin{eqnarray} 
{\sf g}_{\bar{u}\bar{u}} &=& - 1 - \bar{r}^2 {\cal E}_{ab}
\bar{\Omega}^a \bar{\Omega}^b + O(\bar{r}^3/{\cal R}^3) 
\nonumber \\ & & \mbox{} 
+ \frac{2m}{\bar{r}} + 4m \bar{r} {\cal E}_{ab}
\bar{\Omega}^a \bar{\Omega}^b
\nonumber \\ & & \mbox{} 
- 2 \bar{r} \biggl( a_a -\frac{1}{2} h^{\rm tail}_{uua} 
+ h^{\rm tail}_{uau} \biggr) \bar{\Omega}^a
\nonumber \\ & & \mbox{} 
+ O(m \bar{r}^2/{\cal R}^3).  
\label{13} 
\end{eqnarray} 
Comparison with Eq.~(\ref{11}) --- linearized with respect to
$m/\bar{r}$ --- reveals that the acceleration vector must be given by 
\begin{equation} 
a_a = \frac{1}{2} h^{\rm tail}_{uua} - h^{\rm tail}_{uau}.
\label{14}
\end{equation}
As expected, the matching of the perturbed metrics in the buffer 
zone has returned the black hole's equations of motion.  

\subsection*{MiSaTaQuWa equations} 

The tensorial form of Eq.~(\ref{14}) is 
\begin{equation} 
\frac{D u^\mu}{d\tau} = -\frac{1}{2} \bigl( g^{\mu\nu} + u^\mu  
u^\nu \bigr) \bigl( 2 h^{\rm tail}_{\nu\lambda\rho} 
- h^{\rm tail}_{\lambda\rho\nu} \bigr) u^\lambda u^\rho, 
\label{15}
\end{equation} 
and these are the MiSaTaQuWa equations of motion. Here, $z^\mu(\tau)$
gives the description of the black hole's motion in the background
spacetime, $u^\mu = d z^\mu/d\tau$ is the normalized velocity vector,
and $D u^\mu/d\tau$ is the covariant acceleration. All tensors, and
all tensorial operations, refer to the background spacetime of the
external universe; its metric is $g_{\alpha\beta}$, as it was
introduced in Eq.~(\ref{2}). 

The tail field on the right-hand side of Eq.~(\ref{15}) is obtained by
differentiating Eq.~(\ref{9}); when expressed in terms of the original
retarded Green's function of Eq.~(\ref{7}) it is given by 
\begin{eqnarray} 
h^{\rm tail}_{\mu\nu\lambda} &=& 4 m \int_{-\infty}^{\tau-\epsilon}  
\nabla_\lambda \biggl( G_{\mu\nu\mu'\nu'}
- \frac{1}{2} g_{\mu\nu} G^{\rho}_{\ \rho\mu'\nu'}
\biggr) (z, z') 
\nonumber \\ & & \mbox{} 
\times u^{\mu'} u^{\nu'}\, d\tau'. 
\label{16}
\end{eqnarray} 
It is understood that one takes the limit $\epsilon \to 0^+$ of this 
expression. Cutting the integration short excludes the (singular)
light-cone part of the Green's function and isolates its tail part;
the result is a smooth field on the world line. In Eq.~(\ref{16}),
$z \equiv z(\tau)$ stands for the current position on the world line
(at which the self-force is being evaluated) and $z' \equiv z(\tau')$
stands for a prior position. 

\section{Motion of a point mass} 
\label{iv}

The derivation of the MiSaTaQuWa equations sketched in the preceding
section is conceptually sound and devoid of any questionable
assumptions. It is, however, technically involved and one wonders
whether there may not be a shortcut to the final answer. Since
the equations of motion refer to a body without internal structure,
could they be derived on the assumption that the object is a point 
particle? The answer is in the affirmative, provided that one is
willing to introduce additional assumptions and tolerate fields that
are singular on the world line. The derivation sketched below was
first presented by Mino, Tanaka, and Sasaki \cite{mino-etal:97} and
then by Quinn and Wald \cite{quinn-wald:97}; the approach described
here relies heavily on concepts and techniques introduced by Detweiler
and Whiting \cite{detweiler-whiting:03}.  

As we have seen in Sec.~\ref{iii}, the gravitational perturbation
produced by a point particle of mass $m$ is obtained from the
potentials of Eq.~(\ref{8}) by trace reversal: $h_{\alpha\beta} =
\psi_{\alpha\beta} - \frac{1}{2} g_{\alpha\beta} (g^{\gamma\delta}
\psi_{\gamma\delta})$. In Sec.~\ref{iii}, Eq.~(\ref{8}) was assumed to
hold in the external zone only ($r > r_e \gg m$), but we now accept
its global validity. We also accept the fact that the perturbation is
singular on the world line ($r=0$). And we choose not to be bothered
with the fact that while Eq.~(\ref{8}) was obtained by linearizing the
Einstein field equations about the background metric
$g_{\alpha\beta}$, the perturbation is obviously not small when $r$ is
smaller than $m$.  

Our first additional assumption is that the particle will move on a
geodesic of the metric ${\sf g}_{\alpha\beta} = g_{\alpha\beta} +
h_{\alpha\beta}$. Writing down the geodesic equation in terms of
tensorial quantities that refer to the background spacetime (such as
the velocity vector $u^\mu = dz^\mu/d\tau$, which is normalized in the
metric $g_{\alpha\beta}$) returns the equations of motion 
\begin{equation}
\frac{D u^\mu}{d\tau} = -\frac{1}{2} \bigl( g^{\mu\nu} + u^\mu 
u^\nu \bigr) \bigl( 2 h_{\nu\lambda;\rho} - h_{\lambda\rho;\nu} \bigr)
u^\lambda u^\rho.  
\label{17}
\end{equation} 
These are superficially similar to Eq.~(\ref{15}), but the difference
is important. While Eq.~(\ref{15}) involves tensor fields that are
smooth on the world line, Eq.~(\ref{17}) involves highly singular
quantities. These equations of motion are therefore meaningless as
they stand, and $h_{\alpha\beta}$ must be regularized before
Eq.~(\ref{17}) can be evaluated. 

The regularization of a retarded field near its pointlike source is a 
problem that has a long history. This issue was encountered by
Dirac \cite{dirac:38} in the context of the electrodynamics of a point
electric charge in flat spacetime. By appealing to energy-momentum
conservation across a world tube surrounding the charge's world line,
Dirac discovered that regularization could be achieved by decomposing
the electromagnetic field into singular-symmetric ``S'' and
regular-radiative ``R'' pieces; the ``S'' field would simply be
removed from the retarded field and only the remainder --- the ``R''
field --- would be allowed to act on the particle. Dirac further
discovered that in flat spacetime, the ``S'' field is given 
by the following combination of retarded and advanced solutions to
Maxwell's equations: $F^{\rm S}_{\alpha\beta} = \frac{1}{2}
(F^{\rm ret}_{\alpha\beta} + F^{\rm adv}_{\alpha\beta})$. The ``R''
field, on the other hand, is given by $F^{\rm R}_{\alpha\beta} = 
\frac{1}{2}(F^{\rm ret}_{\alpha\beta} 
- F^{\rm adv}_{\alpha\beta})$. Because the ``S'' field satisfies the
same field equations as the retarded field (with a singular source
term on the right-hand side), it is just as singular as the retarded 
field on the world line; removing it from the retarded field produces
a smooth field. This is confirmed by the fact that the ``R'' field
satisfies the sourcefree Maxwell equations. 

Dirac's analysis was generalized to curved spacetime by DeWitt and
Brehme \cite{dewitt-brehme:60}, but the proper decomposition of the
retarded electromagnetic field into ``S'' and ``R'' parts was
identified only recently by Detweiler and Whiting
\cite{detweiler-whiting:03}. For reasons of causality, the
flat-spacetime prescription (a superposition of half retarded and half
advanced fields) does not work in curved spacetime: The retarded field 
depends on the particle's entire past history, the advanced field
depends on its future history, and a linear superposition would depend
on the full history, both past and future (see Fig.~1). This would
give rise to equations of motion with unacceptable causal properties. 

\begin{figure}
\includegraphics[angle=-90,scale=0.33]{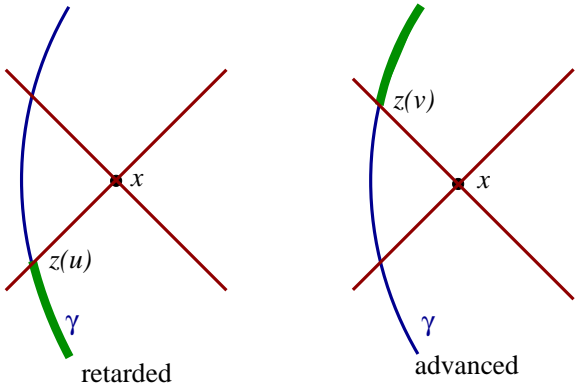}
\caption{Causal properties of the retarded and advanced solutions. The
retarded field at a point $x$ depends on the particle's history prior
to (and including) the retarded point $z(u)$. The advanced solution
depends on the particle's history after (and including) the advanced
point $z(v)$.}
\end{figure}

The correct curved-spacetime prescription (identified by Detweiler and
Whiting, and applied to the gravitational case) is to remove the ``S''
potential 
\begin{eqnarray} 
\psi_{\rm S}^{\alpha\beta}(x) &=& \frac{2m}{r} 
U^{\alpha\beta}_{\ \ \gamma'\delta'}(x,x') u^{\gamma'} u^{\delta'} 
\nonumber \\ & & \mbox{} 
+ \frac{2m}{r} U^{\alpha\beta}_{\ \ \gamma''\delta''}(x,x'') 
u^{\gamma''} u^{\delta''} 
\nonumber \\ & & \mbox{} 
- 2m \int_u^v V^{\alpha\beta}_{\ \ \mu\nu}(x,z) u^{\mu} u^{\nu}\,
d\tau 
\label{18}
\end{eqnarray} 
from the retarded potential. Here $x' \equiv z(u)$ stands for the
retarded point on the world line associated with the field point $x$
(and tensors with primed indices are evaluated at that point), and
$x'' \equiv z(v)$ stands for the advanced point (with a similar
meaning for doubly-primed indices); within the integral $z \equiv
z(\tau)$ stands for an arbitrary point on the world line, and the
integral extends from the retarded point to the advanced
point. Detweiler and Whiting have shown that 
$\psi_{\rm S}^{\alpha\beta}(x)$ satisfies  Eq.~(\ref{6}) --- the same
wave equation as the retarded potential --- and comparison between
Eqs.~(\ref{8}) and (\ref{18}) reveals that the ``S'' potential is just
as singular as the retarded potential on the world line. The ``R''
potential is then defined by 
\begin{equation} 
\psi_{\rm R}^{\alpha\beta}(x) = \psi^{\alpha\beta}(x) 
- \psi_{\rm S}^{\alpha\beta}(x), 
\label{19}
\end{equation}
and it satisfies the homogeneous form of Eq.~(\ref{6}); this is smooth
tensor field on the world line. The causal properties of the ``S'' and
``R'' fields are illustrated in Fig.~2. 

\begin{figure}
\includegraphics[angle=-90,scale=0.33]{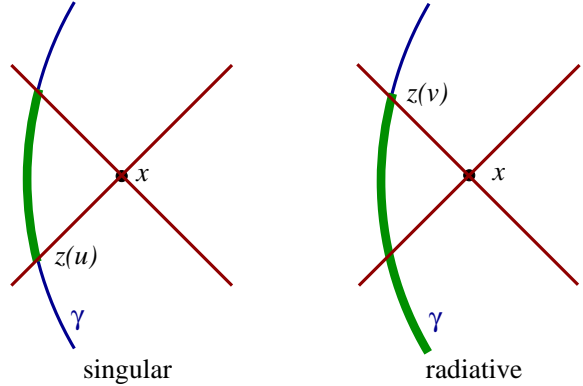}
\caption{Causal properties of the ``S'' and ``R'' fields. The
``S'' field at a point $x$ depends on the particle's history between 
(and including) the retarded point $z(u)$ and the advanced point
$z(v)$. The ``R'' field depends on the particle's history up to the
advanced point $z(v)$. When $x$ is brought to the world line the ``R''
field stays smooth and $z(v)$ coincides with $x$; the field then
depends on the particle's past history only.}  
\end{figure}
 
Having introduced this unique decomposition of the retarded field into
``S'' and ``R'' pieces, Detweiler and Whiting postulate that {\it the
``S'' field exerts no force on the particle; the entire self-force
arises from the action of the ``R'' field}. This axiom can be
seen as another unavoidable assumption that must be introduced in
order to make sense of the motion of point particles. Alternatively,
it can be motivated by showing that the average of the ``S'' field on
a spherical surface (as seen in the particle's instantaneous rest
frame) surrounding the particle is zero. (This calculation is
performed in LRR \cite{poisson:04b}.) In any event, the postulated
equations of motion are     
\begin{equation}
\frac{D u^\mu}{d\tau} = -\frac{1}{2} \bigl( g^{\mu\nu} + u^\mu 
u^\nu \bigr) \bigl( 2 h^{\rm R}_{\nu\lambda;\rho} 
- h^{\rm R}_{\lambda\rho;\nu} \bigr) u^\lambda u^\rho, 
\label{20}
\end{equation} 
and one finds that the ``R'' field reduces to 
\begin{equation} 
h^{\rm R}_{\mu\nu;\lambda} = -4 m \Bigl( u_{(\mu}
R_{\nu)\rho\lambda\xi} + R_{\mu\rho\nu\xi} u_\lambda \Bigr) u^\rho
u^\xi + h^{\rm tail}_{\mu\nu\lambda}         
\label{21}
\end{equation}
on the world line, where $h^{\rm tail}_{\mu\nu\lambda}$ is the tail
field of Eq.~(\ref{16}). Substituting Eq.~(\ref{21}) into
Eq.~(\ref{20}) eliminates the terms involving the Riemann tensor, and
we end up with the MiSaTaQuWa equations of Eq.~(\ref{15}). We conclude
that the equations of motion can indeed be derived on the basis of a
point mass, at the price of introducing the geodesic postulate (which
was not needed for a black hole) and the Detweiler-Whiting axiom
(which also was not needed). 

Equation (\ref{20}) comes with a compelling interpretation. It states
that the point particle moves on a geodesic of the metric
$g_{\alpha\beta} + h^{\rm R}_{\alpha\beta}$, which is smooth on the
world line. Furthermore, because the ``R'' potential satisfies the 
homogeneous form of Eq.~(\ref{6}), this metric is everywhere a vacuum
solution to the Einstein field equations. (Recall that
$g_{\alpha\beta}$ is itself a vacuum solution.) We therefore have a
point mass moving on a geodesic of a well-defined vacuum spacetime.  

\section{Newtonian self-force} 
\label{v}

The notion of a self-force exists also in the simple setting of
Newtonian theory. And the Newtonian potential also can be decomposed
into ``S'' and ``R'' pieces. This pedagogical illustration was first
presented to me by Steve Detweiler. The following presentation draws
heavily from Ref.~\cite{detweiler-poisson:04}.  

Consider, in Newtonian theory, a large mass $M$ at position
$\bm{\rho}(t)$ relative to the centre of mass, and a small mass $m$ at
position $\bm{R}(t)$. We assume that $m \ll M$ and the centre of mass
condition reads $m \bm{R} + M \bm{\rho} = \bm{0}$. We denote the
position of an arbitrary field point by $\bm{x}$, and $r \equiv
|\bm{x}|$ is its distance from the centre of mass. We shall also let
$R \equiv |\bm{R}|$ and $\rho \equiv |\bm{\rho}|$. 

We begin with a test-mass description of the situation, according to 
which the smaller mass moves in the gravitational field of the larger
mass, which is placed at the origin of the coordinate system. The
background Newtonian potential is
\begin{equation}
\Phi_0(\bm{x}) = -\frac{M}{r}
\label{22}
\end{equation}
and the background gravitational field is $\bm{g}_0 = -\bm{\nabla}
\Phi_0 = - M \bm{x} /r^3$. In this description, the smaller mass $m$
moves according to $d^2\bm{R}/dt^2 = \bm{g}_0(\bm{x}=\bm{R})$. If the
motion is circular, then $m$ possesses a uniform angular velocity
given by ${\Omega_0}^2 = M/R^3$, where $R$ is the orbital
radius. These results are in close analogy with a relativistic
description in which the smaller mass is taken to move on a geodesic
of the background spacetime, in a test-mass approximation.

We next improve our description by incorporating the gravitational 
effects produced by the smaller mass. The exact Newtonian potential is
\begin{equation}
\Phi(\bm{x}) = -\frac{M}{|\bm{x} - \bm{\rho}|}
- \frac{m}{|\bm{x} - \bm{R}|},
\label{23}
\end{equation}
and for $m \ll M$ this can be expressed as $\Phi(\bm{x}) =
\Phi_0(\bm{x}) + \delta \Phi(\bm{x})$, with a perturbation given by
\begin{equation}
\delta \Phi(\bm{x}) = -\frac{M}{|\bm{x} - \bm{\rho}|} + \frac{M}{r}
- \frac{m}{|\bm{x} - \bm{R}|}.
\label{24}
\end{equation}
This gives rise to a field perturbation $\delta \bm{g}$ that exerts a force
on the smaller mass. This is the particle's ``bare''
self-acceleration, and the correspondence with the relativistic
problem is clear. 

An examination of Eq.~(\ref{24}) reveals that the last term on the 
right-hand side diverges at the position of the smaller mass. But
since the gravitational field produced by this term is isotropic
around $\bm{R}(t)$, we know that this field will exert no force on
the particle. We conclude that the last term can be identified with
the singular ``S'' part of the perturbation,
\begin{equation}
\Phi_{\rm S}(\bm{x}) = - \frac{m}{|\bm{x} - \bm{R}|},
\label{25}
\end{equation}
and that the remainder makes up the regular ``R'' potential,
\begin{equation}
\Phi_{\rm R}(\bm{x}) = -\frac{M}{|\bm{x} - \bm{\rho}|}
+ \frac{M}{r}.
\label{26}
\end{equation}
The full perturbation is then given by $\delta \Phi(\bm{x}) =
\Phi_{\rm S}(\bm{x}) + \Phi_{\rm R}(\bm{x})$, and only the ``R'' 
potential affects the motion of the smaller mass. Once more the
correspondence with the relativistic problem is clear.

It is easy to check that to first order in $m/M$, Eq.~(\ref{26})
simplifies to
\begin{equation}
\Phi_{\rm R}(\bm{x}) = m \frac{ \bm{R} \cdot \bm{x} }{r^3};
\label{27}
\end{equation}
this simplification occurs because of the centre-of-mass
condition, which implies that $\bm{\rho}$ is formally of order 
$m/M \ll 1$. The ``R'' part of the field perturbation is then 
\begin{equation}
\bm{g}_{\rm R}(\bm{x}) = m \frac{ 3 (\bm{R} \cdot \bm{x}) \bm{x}
- r^2 \bm{R}}{r^5},
\label{28}
\end{equation}
and evaluating this at the particle's position yields a correction to
the background field $\bm{g}_0(\bm{x}=\bm{R}) = -M \bm{R}/R^3$ given
by $\bm{g}_{\rm R} (\bm{x}=\bm{R}) = 2m \bm{R}/R^3$; the force still
points in the radial direction but the active mass has been shifted
from $M$ to $M-2m$. For circular motion the angular velocity becomes
$\Omega^2 = (M-2m)/R^3$. This can be cast in a more recognizable form
if we express the angular velocity in terms of the total separation $s
\equiv R+\rho = (1+m/M)R$ between the two masses. To first order in
$m/M$ we obtain $\Omega^2 = (M+m)/s^3$, which is just the usual form
of Kepler's third law. The ``R'' part of the field perturbation is
therefore responsible for the finite-mass correction to the angular
velocity. 

Notice that the ``R'' potential of Eq.~(\ref{27}) has a pure dipolar
form. By contrast, the potentials $\delta \Phi$ and $\Phi_{\rm S}$
contain an infinite number of multipole moments. This observation
should be kept in mind as the reader proceeds through the remaining
sections of this contribution.    
 
\section{Concrete evaluation of the self-force} 
\label{vi}

I turn next to the practical issues involved in a concrete evaluation 
of the gravitational self-force --- the right-hand side of
Eq.~(\ref{20}), which shall be denoted $a^\mu[h^{\rm R}]$ --- for a
particle moving in the field of a Schwarzschild or Kerr black hole.  

The first sequence of steps are concerned with the computation of the
(retarded) metric perturbation $h_{\alpha\beta}$ produced by a point 
particle moving on a specified geodesic of the Kerr spacetime. A
method for doing this was elaborated by Lousto and Whiting
\cite{lousto-whiting:02} and Ori \cite{ori:03}, building on the
pioneering work of Teukolsky \cite{teukolsky:73}, Chrzanowski
\cite{chrzanowski:75}, and Wald \cite{wald:78}. The procedure consists
of (i) solving the Teukolsky equation for one of the Newman-Penrose
quantities $\psi_0$ and $\psi_4$ (which are complex components of the
Weyl tensor) produced by the point particle; (ii) obtaining from
$\psi_0$ or $\psi_4$ a related (Hertz) potential $\Psi$ by integrating
an ordinary differential equation; (iii) applying to $\Psi$ a number
of differential operators to obtain the metric perturbation in a 
radiation gauge that differs from the Lorenz gauge of Eq.~(\ref{5});
and (iv) performing a gauge transformation from the radiation gauge to 
the Lorenz gauge. For a Schwarzschild black hole one can rely instead
on the formalism of metric perturbations \cite{regge-wheeler:57,
zerilli:70}, and the procedure simplifies.  

It is well known that the Teukolsky equation separates when
$\psi_0$ or $\psi_4$ is expressed as a multipole expansion, summing
over modes with (spheroidal-harmonic) indices $l$ and $m$. In fact,
the procedure outlined above relies heavily on this mode
decomposition, and the metric perturbation returned at the end of the
procedure is also expressed as a sum over modes:
\begin{equation}
h_{\alpha\beta} = \sum_\ell h^\ell_{\alpha\beta}. 
\label{29}
\end{equation} 
(For each $l$, $m$ ranges from $-l$ to $l$, and summation of $m$ over
this range is henceforth understood. Indices on the metric
perturbation and other tensors will now be omitted for ease of
notation.) From the modes $h_\ell$, mode contributions to the
self-acceleration can be computed: $a[h_\ell]$ is obtained from
Eq.~(\ref{20}) by substituting $h_\ell$ in place of $h^{\rm R}$. These
mode contributions are finite on the world line, but $a[h_\ell]$
is discontinuous at the radial position of the particle. The sum over
modes does not converge, because the ``bare'' acceleration
(constructed from the retarded field $h$) is formally infinite.  

The next sequence of steps is concerned with the regularization of 
each $a[h_l]$ by removing the contribution from 
$h^{\rm S}$. Starting with Eq.~(\ref{18}), the singular field can be
constructed locally in a neighbourhood of the particle, and then
decomposed into modes of multipole order $\ell$. This gives rise to
modes $a[h^{\rm S}_\ell]$ for the singular part of the
self-acceleration; these are also finite and discontinuous, and their
sum over $\ell$ also diverges. But the true modes $a[h^{\rm R}_\ell] 
= a[h_\ell] - a[h^{\rm S}_\ell]$ of the self-acceleration are
continuous at the radial position of the particle, and their sum does
converge to the particle's acceleration. 

The general structure of $a[h^{\rm S}_\ell]$, as worked out in
Refs.~\cite{barack-etal:02, barack-ori:02, barack-ori:03a, 
barack-ori:03b, mino-etal:03, detweiler-etal:03, kim:04}, is given by 
\begin{equation} 
a^\mu[h^{\rm S}_\ell] = (\ell + {\textstyle \frac{1}{2}}) A^\mu +
B^\mu + \frac{C^\mu}{\ell + {\textstyle \frac{1}{2}}} 
+ \mbox{convergent terms}. 
\label{30}
\end{equation} 
The ``regularization parameters'' $A^\mu$, $B^\mu$, and $C^\mu$ are
independent of $\ell$ but they depend on the details of the spacetime
and on the particle's state of motion. It is evident that the term
involving $A^\mu$ in Eq.~(\ref{30}) will diverge quadratically when
summed over $\ell$, that the term involving $B^\mu$ will diverge
linearly, and that the term involving $C^\mu$ will diverge
logarithmically; all other terms converge, and keeping additional
terms produces faster convergence of the sum $\sum_{\ell} 
a[h^{\rm R}_\ell]$. 

The structure of Eq.~(\ref{30}) can easily be recovered in the  
Newtonian setting of the preceding section. The multipolar
decomposition of the singular potential of Eq.~(\ref{25}) is given by 
\begin{equation} 
\Phi_{\rm S}(\bm{x}) = - m \sum_{\ell}
\frac{(r_<)^\ell}{(r_>)^{\ell+1}} P_\ell( \bm{\hat{n}} \cdot
\bm{\hat{R}} ), 
\label{31}
\end{equation}
where $r_< = \mbox{min}(r,R)$, $r_> = \mbox{max}(r,R)$, $\bm{\hat{n}}
= \bm{x}/r$, $\bm{\hat{R}} = \bm{R}/R$, and $P_\ell(\mu)$ are Legendre
polynomials. Taking the gradient of Eq.~(\ref{31}) and then the limit
$\bm{x} \to \bm{R}$ returns 
\begin{equation} 
g^r_{\rm S} = \frac{m}{R^2} \sum_{\ell} \biggl[ \mp (\ell  
+ {\textstyle \frac{1}{2}}) - \frac{1}{2} \biggr]
\label{32}
\end{equation}
for the radial component of the singular field (the angular components
all vanish). This has the same form as Eq.~(\ref{30}), with $A^r = \mp
m/R^2$, $B^r = -m/(2R^2)$, and $C^r = 0$. The choice of sign in front
of the $(\ell + {\textstyle \frac{1}{2}})$ term in Eq.~(\ref{32})
comes from the two ways in which the limit $r \to R$ can be taken: The
upper (minus) sign corresponds to $r \to R^+$ while the lower (plus)
sign corresponds to $r \to R^-$. This Newtonian calculation reproduces
the leading-order term in a post-Newtonian expansion of the
relativistic regularization parameters \cite{barack-etal:02,
barack-ori:02, barack-ori:03a, barack-ori:03b, mino-etal:03,
detweiler-etal:03, kim:04}.   

The self-acceleration is obtained by first computing $a[h_\ell]$
from the retarded metric perturbation, then computing the counterterms 
$a[h^{\rm S}_\ell]$ by mode-decomposing the singular field, and
finally summing over all $a[h^{\rm R}_\ell] = a[h_\ell]
- a[h^{\rm S}_\ell]$. This procedure is lengthy and involved, and 
thus far it has not been brought to completion, except for the special
case of a particle falling radially toward a nonrotating black hole
\cite{barack-lousto:02}. Another evaluation of the self-force, which
did not rely on a mode decomposition, was carried out in the context
of weak fields and slow motions \cite{pfenning-poisson:02}; this
reproduced the standard post-Newtonian result.  

The procedure described above is lengthy and involved, but it is also
incomplete when the background spacetime is that of a Kerr black
hole. The reason is that the metric perturbations
$h^\ell_{\alpha\beta}$ that can be recovered from $\psi_0$ or $\psi_4$
do not by themselves sum up to the complete gravitational perturbation 
produced by the moving particle. Missing are the perturbations derived
from the other Newman-Penrose quantities: $\psi_1$, $\psi_2$, and
$\psi_3$. While $\psi_1$ and $\psi_3$ can always be set to zero by an
appropriate choice of null tetrad, $\psi_2$ contains such important
physical information as the shifts in mass and angular-momentum
parameters produced by the particle \cite{wald:73}. Because the mode 
decompositions of $\psi_0$ and $\psi_4$ start at $l=2$, one might
say that the ``$\ell=0$ and $\ell=1$'' modes of the metric
perturbations are missing. It is not  
currently known how the procedure can be completed so as to
incorporate {\it all modes} of the metric perturbations. Specializing
to a Schwarzschild spacetime eliminates this difficulty, and in this
context the low multipole modes have been studied for the special case
of circular orbits \cite{nakano-etal:03, detweiler-poisson:04}. In
view of the fact that the Newtonian self-force is purely $\ell = 1$
(refer back to the last paragraph of Sec.~\ref{v}), it is clear that
these low multipoles cannot be ignored. 

Taking into account these many difficulties (and I choose to stay
silent on others, for example, the issue of relating metric
perturbations in different gauges when the gauge transformation is
singular on the world line), it is perhaps not too surprising that
such a small number of concrete calculations have been presented to
date. But progress in dealing with these difficulties has been steady,
and the situation should change dramatically in the next few years. 

\section{Beyond the self-force} 
\label{vii}

The successful computation of the gravitational self-force is not the  
end of the story. After the difficulties reviewed in the preceding
section have all been dealt with and the motion of the small body 
is finally calculated to order $m$, it will still be necessary to
obtain gauge-invariant information associated with the body's
corrected motion. Because the MiSaTaQuWa equations are not
by themselves gauge-invariant [they are formulated in the Lorenz gauge
of Eq.~(\ref{5})], this step will necessitate going beyond the
self-force.  

To see how this might be done, imagine that the small body is a 
pulsar, and that it emits light pulses at regular proper-time
intervals. The motion of the pulsar around the central black hole 
modulates the pulse frequencies as measured at infinity, and
information about the body's corrected motion is encoded in
the times-of-arrival of the pulses. Because these can be measured
directly by a distant observer, they clearly constitute
gauge-invariant information. But the times-of-arrival are determined
not only by the pulsar's motion, but also by the propagation of
radiation in the perturbed spacetime. This example shows that to
obtain gauge-invariant information, one must properly combine the
MiSaTaQuWa equations of motion with the metric perturbation.   

In the astrophysical context of the Laser Interferometer Space
Antenna, reviewed in Sec.~\ref{ii}, the relevant observable is the
instrument's response to a gravitational wave, which is determined by
gauge-invariant waveforms, $h_+$ and $h_\times$. To calculate these is
the ultimate goal of the Capra program, and the challenges that lie
ahead go well beyond what I have described thus far. To obtain the
waveforms it will be necessary to solve the Einstein field equations
to {\it second order} in perturbation theory.   

To understand this, consider first the formulation of the first-order
problem. Schematically, one introduces a perturbation $h$ that
satisfies a wave equation $\Box h = T[z]$ in the background
spacetime. Here $T[z]$ is the stress-energy tensor of the moving 
body, which is a functional of the world line $z(\tau)$. In
first-order perturbation theory, the stress-energy tensor must be
conserved in the background spacetime, and $z(\tau)$ must describe a
geodesic. It follows that in first-order perturbation theory, the
waveforms constructed from the perturbation $h$ contain no information
about the body's corrected motion.

The first-order perturbation, however, can be used to correct the
motion, which is now described by the world line $z(\tau) + \delta
z(\tau)$. In a naive implementation of the self-force, one would now
re-solve the wave equation with a corrected stress-energy tensor,
$\Box h = T[z + \delta z]$, and the new waveforms constructed from
$h$ would then incorporate information about the corrected
motion. This implementation is naive because this information would
not be gauge-invariant. In fact, to be consistent one would have to
include {\it all} second-order terms in the wave equation, not 
just the ones that come from the corrected motion. Schematically, the
new wave equation would have the form of $\Box h = (1 + h) T[z +
\delta z] + (\nabla h)^2$, and this is much more difficult to solve
than the naive problem (if only because the source term is now
much more singular than the distributional singularity contained in
the stress-energy tensor). But provided one can find a way to make
this second-order problem well posed, and provided one can solve it
(or at least the relevant part of it), the waveforms constructed from 
the second-order perturbation $h$ will be gauge invariant. In this
way, information about the body's corrected motion will have properly
been incorporated into the gravitational waveforms.        

\section{Conclusion} 
\label{viii}
  
There has been significant progress toward solving the self-force
problem over the last several years. The foundations (reviewed in
Secs.~\ref{iii} and \ref{iv}) are now solid, thanks to the work of
Mino, Sasaki, and Tanaka \cite{mino-etal:97}, Quinn and Wald
\cite{quinn-wald:97}, and Detweiler and Whiting
\cite{detweiler-whiting:03}. The regularization parameters $A^\mu$,
$B^\mu$, and $C^\mu$ (introduced in Sec.~\ref{vi}) have been
calculated for arbitrary motion in the Schwarzschild and Kerr
spacetimes by Barack and Ori \cite{barack-etal:02, barack-ori:02,
barack-ori:03a, barack-ori:03b} and Mino, Nakano, and Sasaki
\cite{barack-etal:02, mino-etal:03}, and their expressions were 
independently verified \cite{detweiler-etal:03, kim:04}. And finally,
the reconstruction of the metric perturbation from the Teukolsky
function (described in Sec.~\ref{vi}) is now better understood, thanks
to the work of Lousto and Whiting \cite{lousto-whiting:02} and Ori
\cite{ori:03}. 

While this progress is encouraging, the challenges that lie ahead
are still numerous. Among the outstanding issues are (i) the question
of defining and calculating the ``$\ell = 0$ and $\ell = 1$''
contributions to the self-force in the Kerr spacetime (as was
discussed in Sec.~\ref{vi}), (ii) the design of useful
gauge-invariant quantities that could be computed from the self-force
and the metric perturbation (as was discussed in Sec.~\ref{vii}), and
(iii) the incorporation of the self-force into a consistent
wave-generation formalism (as was also discussed in
Sec.~\ref{vii}). These issues are fascinating, and in the next few
years they will be vigourously pursued by the Capra posse.     

\subsection*{Acknowledgments} 

I wish to thank the organizers of GR17 for inviting me to talk about
the gravitational self-force at this prestigious international
conference. And I wish to thank the Capra posse (listed by name at the
end of Sec.~\ref{i}) for numerous discussions on this topic. The work
presented in this contribution was supported by the Natural Sciences
and Engineering Research Council of Canada.      
 
\bibliography{motion}
\end{document}